\documentstyle[epsfig,rotate]{aipproc}

\newcommand{\ba}{\begin{eqnarray}}
\newcommand{\ea}{\end{eqnarray}}
\input{psfig}

\begin{document}
\pagestyle{plain}

\title{Electromagnetic Form Factors of the Nucleon}
\author{R. Bijker$^{1)}$ and A. Leviatan$^{2)}$}
\address{
$^{1)}$ Instituto de Ciencias Nucleares, 
        Universidad Nacional Aut\'onoma de M\'exico, 
        A.P. 70-543, 04510 M\'exico D.F., M\'exico\\
$^{2)}$ Racah Institute of Physics, 
        The Hebrew University, Jerusalem 91904, Israel}

\maketitle

\begin{abstract}
We reanalyze the world data on the electromagnetic form factors 
of the nucleon. The calculations are performed in the framework 
of an algebraic model of the nucleon combined with 
vector meson dominance.
\end{abstract}

\section{Introduction}

The electromagnetic form factors of the nucleon and its 
excitations (baryon resonances) provide a powerful tool 
to investigate the structure of the nucleon \cite{Baryons}. 
These form factors can be measured in electroproduction 
as a function of the four-momentum squared $q^2=-Q^2$ of the virtual 
photon. Especially the transition to the region of high momentum 
transfer, for which the methods of perturbative QCD apply,  
is currently of much interest \cite{Stoler}. 

In this contribution we present an analysis of the world data 
for the elastic form factors $G_{E/M}^{p/n}$.
Our method is a combination of a recently introduced algebraic 
model of the nucleon \cite{BIL} and vector meson 
dominance.

\section{Algebraic model}

The algebraic approach provides a unified treatment of various 
constituent quark models \cite{BIL}, such as harmonic oscillator 
quark models and collective models. In this contribution we employ 
a collective model of the nucleon in which baryon resonances are 
interpreted as vibrational and rotational excitations of an 
oblate top. There are two fundamental vibrations: a breathing 
mode and a two-dimensional vibrational mode, which are 
associated with the $N(1440)P_{11}$ Roper resonance and the 
$N(1710)P_{11}$ resonance, respectively. The negative parity 
resonances of the second resonance region are interpreted as 
rotational excitations. Since each vibrational mode 
has its own characteristic frequency, this collective model has 
no problem with the relative energy of the Roper resonance with 
respect to the negative parity resonances. 

In \cite{emff} we studied the elastic electromagnetic form factors 
of the nucleon. These calculations include anomalous magnetic moments 
for the proton and the neutron, as well as a flavor dependent 
distribution functions of the charge and magnetization. 
Supposedly, the anomalous magnetic moments and the flavor dependence 
arise as effective parameters, since the coupling to the meson cloud 
surrounding the nucleon was not included explicitly. 
According to \cite{emff} the electric and magnetic form factors 
of the nucleon, when folded with a distribution of the charge 
and magnetization, can be expressed in terms of a common intrinsic 
dipole form factor
\ba
G_E^p(Q^2) \;=\; G_M^p(Q^2) \;=\; g(Q^2) \;=\; 1/(1+\gamma Q^2)^2 ~,
\nonumber\\
G_M^n(Q^2)/G_M^p(Q^2) \;=\; -2/3 ~, 
\hspace{1cm} G_E^n(Q^2) \;=\; 0 ~.
\label{Sachs}
\ea
Note that the Sachs form factors of Eq.~(\ref{Sachs}) do not contain 
anomalous magnetic moments nor involve 
flavor dependent distribution functions. 
In order to study the coupling to the meson cloud we express 
the Sachs form factors in terms of the isoscalar and isovector 
Dirac ($F_1^{S/V}$) and Pauli ($F_2^{S/V}$) form factors
\ba
F_1^S(Q^2) \;=\; g(Q^2) \, \frac{1+\frac{1}{3}\tau}{1+\tau} ~,
\hspace{1cm} 
F_1^V(Q^2) \;=\; g(Q^2) \, \frac{1+\frac{5}{3}\tau}{1+\tau} ~,
\nonumber\\
F_2^S(Q^2) \;=\; -F_2^V(Q^2) \;=\; 
-\frac{2}{3} \, g(Q^2) \, \frac{1}{1+\tau} ~, \hspace{1.5cm} 
\ea
with $\tau=Q^2/4M^2$. 

\section{Meson cloud couplings}

The effects of the meson cloud surrounding the nucleon are taken into 
account in a similar way as in \cite{IJL}, {\it i.e.} by including the 
coupling to the isoscalar vector mesons $\omega$ and $\phi$ and the 
isovector vector meson $\rho$. 
These contributions are studied phenomenologically by parametrizing 
the Dirac and Pauli form factors as 
\ba
F_1^S(Q^2) &=& g(Q^2) \, \frac{1+\frac{1}{3}\tau}{1+\tau} 
\left[ \beta^S + \beta_{\omega} \, \frac{m_{\omega}^2}{m_{\omega}^2+Q^2} 
+ \beta_{\phi} \, \frac{m_{\phi}^2}{m_{\phi}^2+Q^2} \right] ~,
\nonumber\\
F_1^V(Q^2) &=& g(Q^2) \, \frac{1+\frac{5}{3}\tau}{1+\tau} 
\left[ \beta^V + \beta_{\rho} \, 
\frac{m_{\rho}^2}{m_{\rho}^2+Q^2} \right] ~,
\nonumber\\
F_2^S(Q^2) &=& -\frac{2}{3} \, g(Q^2) \, \frac{1}{1+\tau} 
\left[ \alpha^S 
+ \alpha_{\omega} \, \frac{m_{\omega}^2}{m_{\omega}^2+Q^2} 
+ \alpha_{\phi} \, \frac{m_{\phi}^2}{m_{\phi}^2+Q^2} \right] ~,
\nonumber\\
F_2^V(Q^2) &=& \frac{2}{3} \, g(Q^2) \, \frac{1}{1+\tau}  
\left[ \alpha^V 
+ \alpha_{\rho} \, \frac{m_{\rho}^2}{m_{\rho}^2+Q^2} \right] ~.
\ea
The large width of the $\rho$ meson ($\Gamma_{\rho}=151$ MeV) is taken 
into account by making the replacement \cite{IJL} 
\ba
\frac{m_{\rho}^2}{m_{\rho}^2+Q^2} &\rightarrow& 
\frac{m_{\rho}^2 + 8 \Gamma_{\rho} m_{\pi}/\pi} 
{m_{\rho}^2+Q^2 + (4m_{\pi}^2+Q^2) \Gamma_{\rho} \alpha(Q^2)/m_{\pi}} 
\ea
with 
\ba
\alpha(Q^2) &=& \frac{2}{\pi} 
\left[ \frac{4m_{\pi}^2+Q^2}{Q^2} \right]^{1/2} 
\ln \left( \frac{\sqrt{4m_{\pi}^2+Q^2} + \sqrt{Q^2}}{2m_{\pi}} \right) ~.
\ea
The coefficients $\beta^{S/V}$ and $\alpha^{S/V}$ are determined by the 
electric charges and the magnetic moments of the nucleon, respectively.

For small values of the momentum transfer the Dirac and Pauli form 
factors are dominated by the meson dynamics and reduce to a monopole 
form, whereas for high values they show the $Q^2$ dependence as 
predicted by perturbative QCD 
\ba
F_1^{S/V} \sim 1/Q^4 ~, \hspace{1cm} 
F_2^{S/V} \sim 1/Q^6 ~.
\ea

\section{Results}

\begin{figure}
\vfill 
\begin{minipage}{.48\linewidth}
\centerline{\epsfig{file=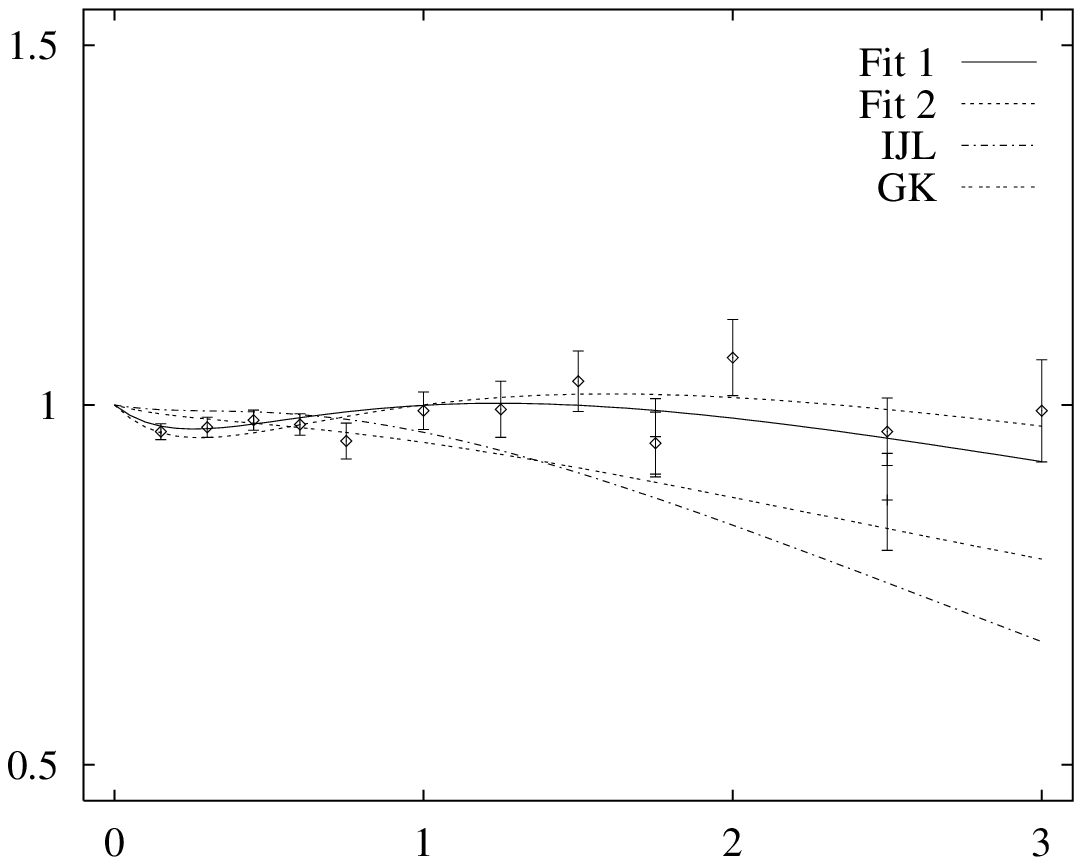,width=0.95\linewidth}}
\caption[]{Proton electric form factor 
$G_E^p/F_D$ as a function of $Q^2$ in (GeV/c)$^2$.}
\label{gepfd}
\end{minipage}\hfill
\begin{minipage}{.48\linewidth}
\centerline{\epsfig{file=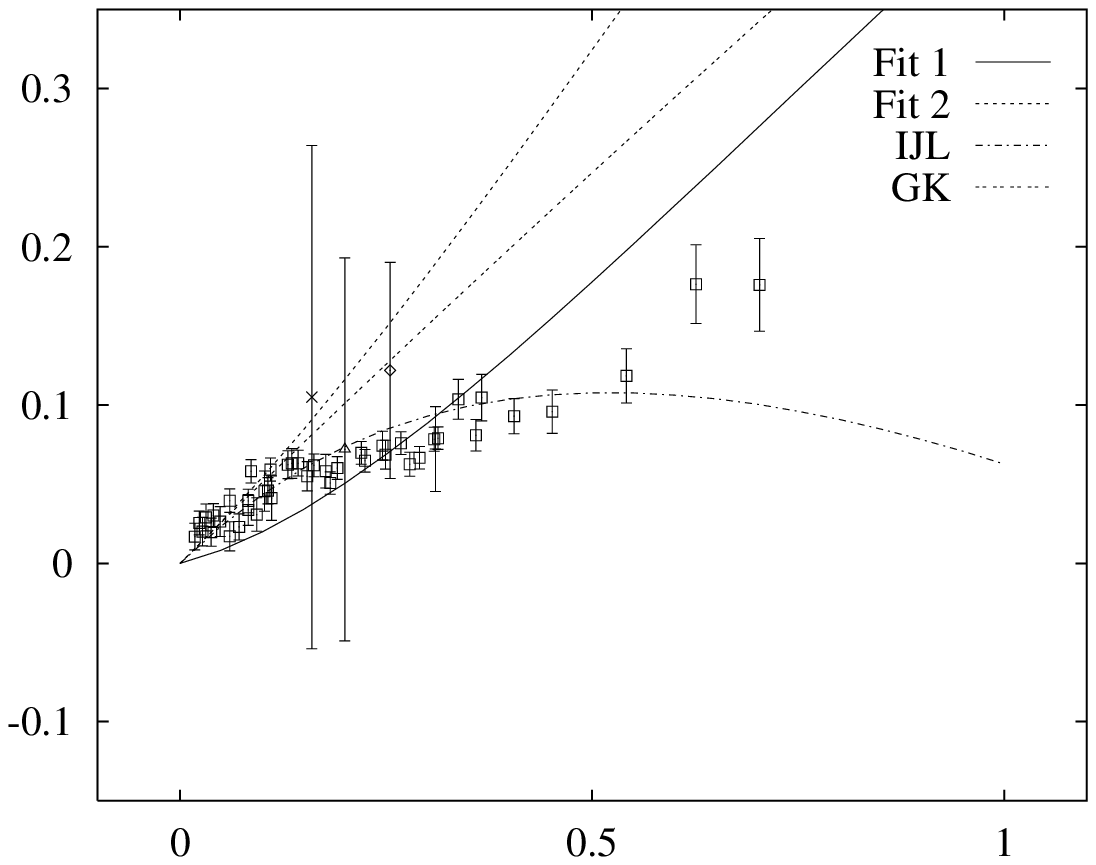,width=0.95\linewidth}}
\caption[]{Neutron electric form factor $G_E^n/F_D$.}
\label{gen}
\end{minipage}
\vspace{15pt}
\begin{minipage}{.48\linewidth}
\centerline{\epsfig{file=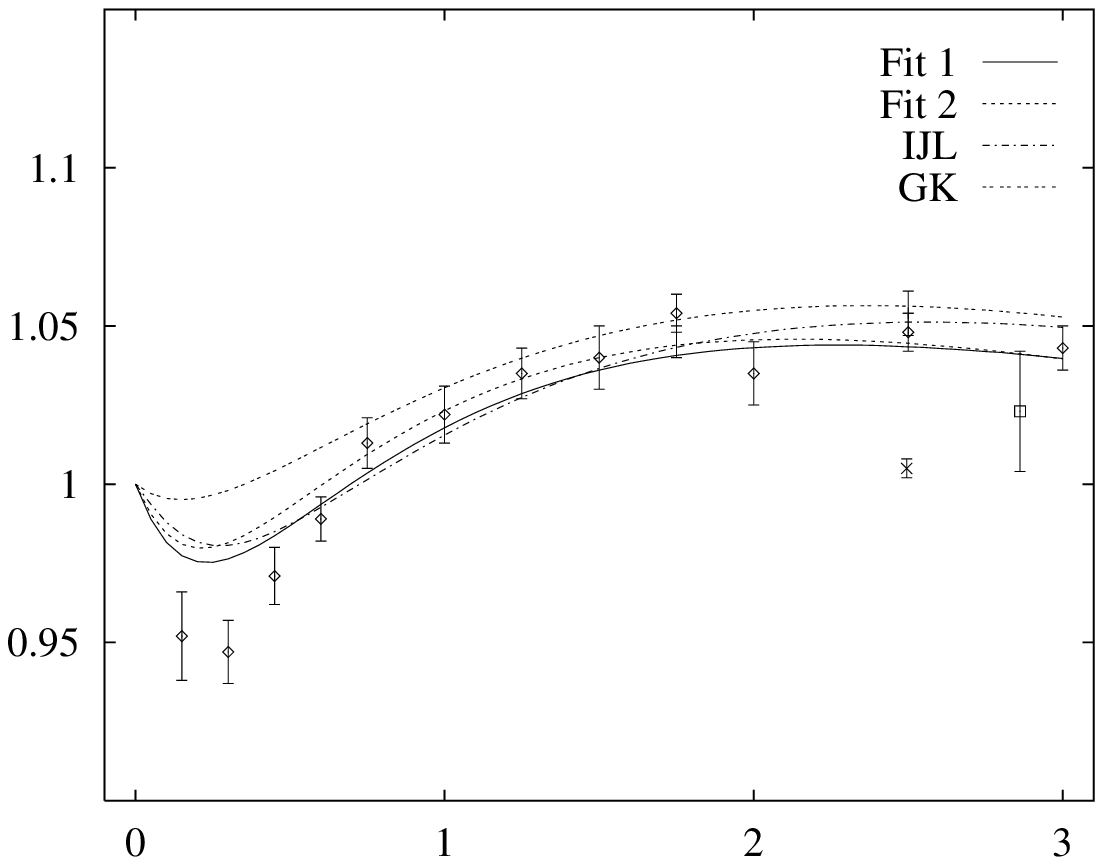,width=0.95\linewidth}}
\caption{Proton magnetic form factor $G_M^p/\mu_p F_D$.}
\label{gmpfd}
\end{minipage}\hfill
\begin{minipage}{.48\linewidth}
\centerline{\epsfig{file=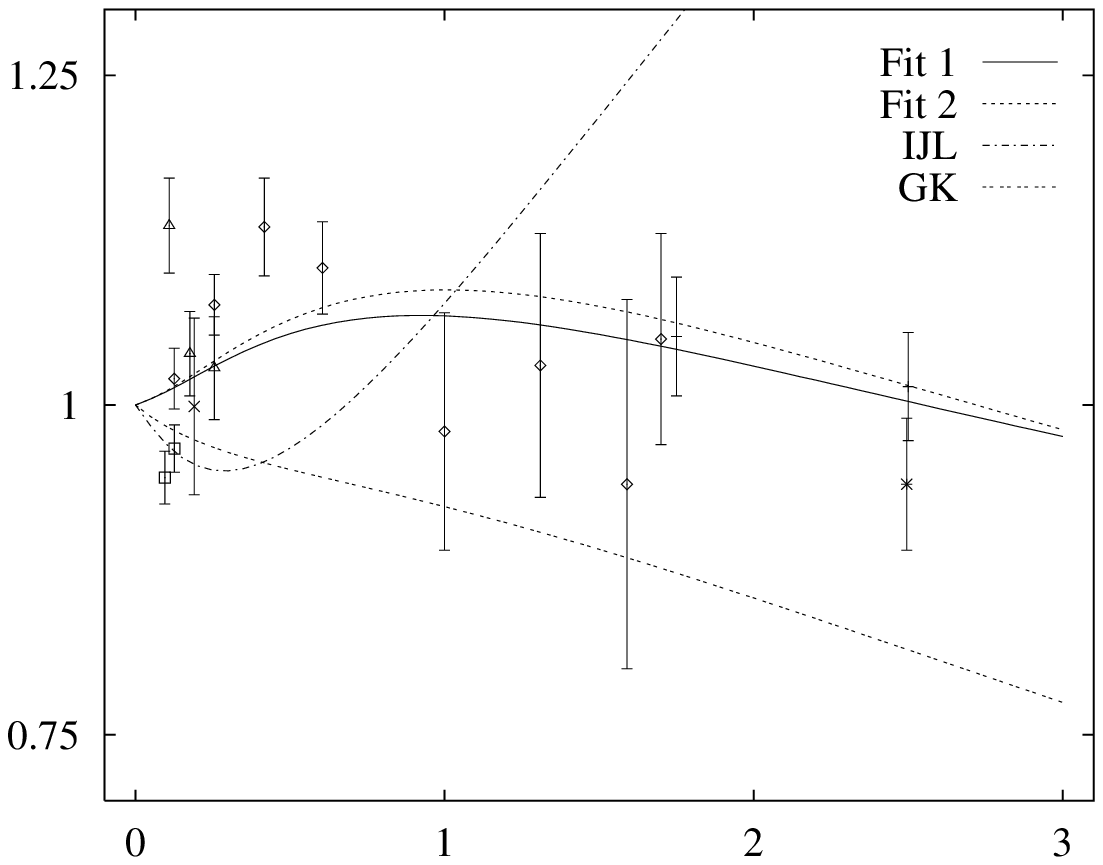,width=.95\linewidth}}
\caption{Neutron magnetic form factor $G_M^n/\mu_n F_D$.}
\label{gmnfd}
\end{minipage}
\vspace{15pt}
\begin{minipage}{.48\linewidth}
\centerline{\epsfig{file=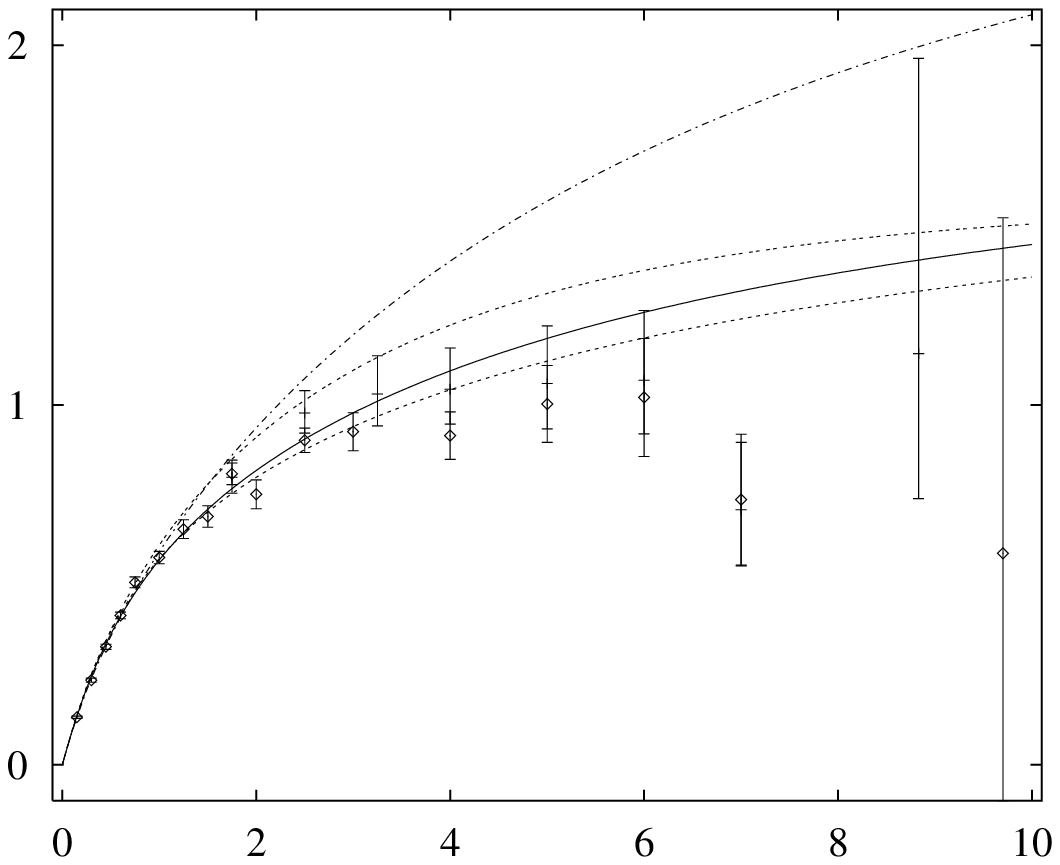,width=0.95\linewidth}}
\caption{Ratio of proton form factors $Q^2 F_2^p/F_1^p$.}
\label{fprat}
\end{minipage}\hfill
\begin{minipage}{.48\linewidth}
\centerline{\epsfig{file=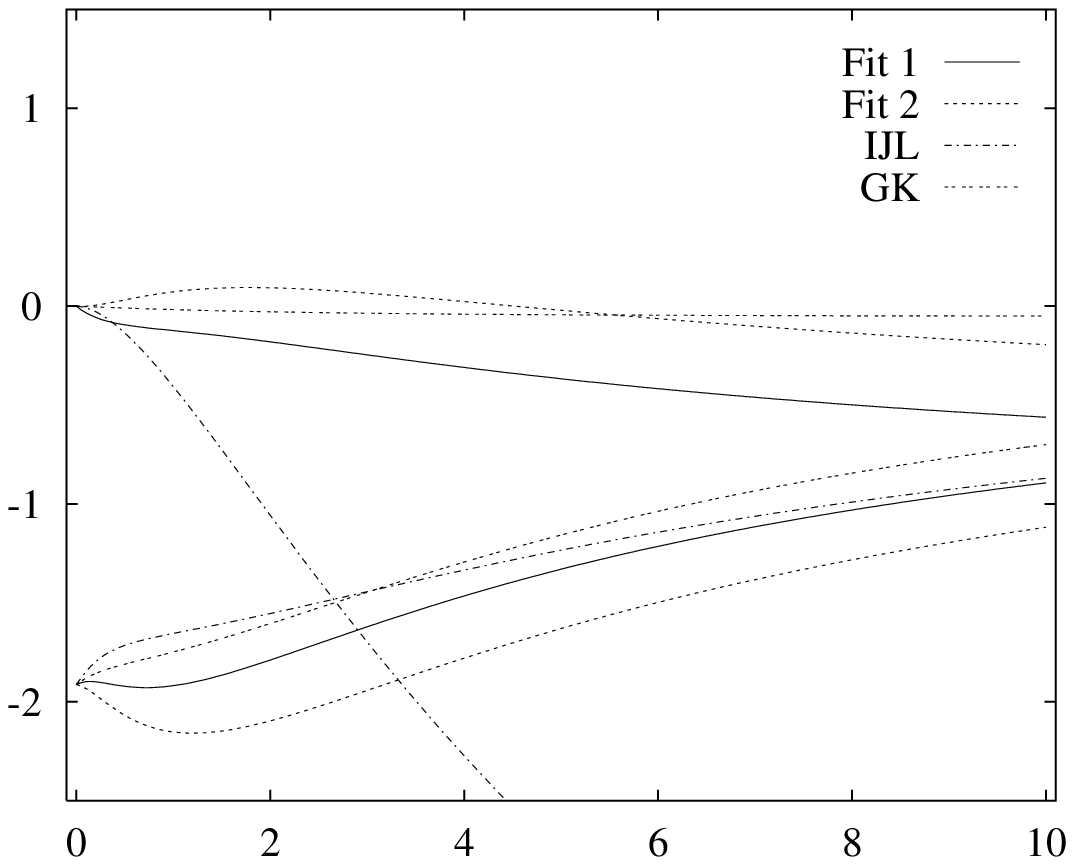,width=.95\linewidth}}
\caption{Neutron form factors $F_1^n/F_D$ and $F_2^n/F_D$.}
\label{fnfd}
\end{minipage}
\end{figure}

Recently, the electromagnetic form factors of the nucleon 
have been remeasured (or reanalyzed) \cite{Bosted,Ostrick}. 
In Figs.~\ref{gepfd}--\ref{gmnfd} we show a compilation of the 
world data on the electromagnetic form factors of the nucleon. 
In the present calculation the coefficients $\beta_M$ and $\alpha_M$ 
(with $M=\rho$, $\omega$, $\phi$) and the scale parameter $\gamma$ 
in the dipole form factor $g(Q^2)$ are determined in a simultaneous 
fit to all four electromagnetic form factors of the nucleon 
with $Q^2 \leq 10$ (GeV/c)$^2$ (solid lines, Fit 1). 
As usual, the form factors are scaled by the standard dipole 
fit $F_D=1/(1+Q^2/0.71)^2$. The oscillations around the dipole 
values are due to the meson cloud couplings. The magnetic form 
factors show an interesting behavior. Whereas $G_M^p$ first decreases 
with respect to the dipole, the new measurements of $G_M^n$ show 
an increase with respect to $F_D$ \cite{Ostrick,Schoch}. This 
behavior is reproduced by the present calculation.  

The electric form factor of the neutron is the least known. 
Unlike for the proton, the Rosenbluth separation of $G_E^n$ from  
$G_M^n$ for a neutron target is difficult for all values of $Q^2$: 
for small $Q^2$ because of the small size of $G_E^n$ compared to 
$G_M^n$, and for large $Q^2$ because the magnetic component 
dominates both the angular dependent and angular independent 
term in the cross section. For this reason we have carried out a 
second fit, in which $G_E^n$ is excluded from the fitting procedure 
and replaced by the proton and neutron charge radii 
(dashed lines, Fit 2). In this calculation the neutron charge radius 
(the slope of $G_E^n$ in the origin) is reproduced, but the existing 
data for $G_E^n$ are overpredicted. The changes for the other 
form factors are minor.  

In Fig.~\ref{fprat} we show the scaling property of the proton 
form factors: $Q^2 F_2^p/F_1^p \sim 1$. The Dirac and Pauli form 
factors of the neutron are presented in Fig.~\ref{fnfd}. Since 
in our calculations the Dirac form factor is small $F_1^n \approx 0$, 
we find $G_E^n \approx -\tau G_M^n$. For $\tau \approx 1$ 
($Q^2 \approx 4M^2$) the electric and magnetic form factor become 
comparable in size. The same effect was pointed out in \cite{GK}. 

For comparison we also show the results of two other calculations: 
the vector meson dominance model of Iachello, Jackson and Lande \cite{IJL}
(dash-dotted lines, IJL), and a hybrid model (interpolation 
between vector meson dominance and pQCD) by Gari and Kr\"umpelmann 
\cite{GK} (dash-dashed lines, GK).

\section{Conclusions}

We have presented a simultaneous analysis of the elastic 
electromagnetic form factors of the nucleon in the context of an 
algebraic model of the nucleon combined with vector meson dominance. 
For a phenomenological approach (as the present one) a good 
data set is a prerequisite.  
Whereas the proton form factors are relatively well known, there is 
still some controversy about the neutron form factors. 
New measurements of the polarization asymmetry in which the ratio of 
the electric and magnetic form factor of the neutron is extracted 
\cite{Ostrick,Burkert} may help to clarify the experimental situation. 

In addition to the elastic form factors discussed in this contribution, 
there is currently much interest in the inelastic transition form 
factors \cite{Stoler}. We plan to extend the present approach to 
include the resonance form factors as well, in order to analyze all 
electromagnetic form factors within the same framework.

\section*{Acknowledgements}
It is a pleasure to thank P.E. Bosted, E.E.W. Bruins and P. Stoler for 
sharing their respective compilations of the world data on the nucleon 
form factors. 
The work is supported in part by grant No. 94-00059 from the United 
States-Israel Binational Science Foundation (BSF), Jerusalem, Israel 
(A.L.) and by DGAPA-UNAM under project IN105194 (R.B.).

\end{document}